\documentclass[aps, llncs, preprint, showkeys]{revtex4-1} 
\usepackage{graphicx}
\usepackage{amsmath}
\usepackage{float}
\usepackage{color}
\usepackage{soul}
\usepackage{xcolor}
\usepackage[dvipsnames]{xcolor}

\begin{document}
\title{Magnetic field decouples nodeless surface and nodal bulk orders in PdTe}

\author{Atanu Mishra$^1$, Ghulam Mohmad$^1$, Kiran Bansal$^1$, Mohd Monish$^1$, Pankaj Kumar$^2$, Chandrasekhar Yadav$^2$, Goutam Sheet$^1$}
%\email{mona@iisermohali.ac.in}
\email{goutam@iisermohali.ac.in}

\affiliation{$^1$Department of Physical Sciences, Indian Institute of Science Education and Research (IISER) Mohali, Sector 81, S. A. S. Nagar, Manauli, PO 140306, India}
\affiliation{$^2$School of Physical Sciences, Indian Institute of Technology Mandi, Kamand 175005, Himachal Pradesh, India}

\begin{abstract}

Selective spectroscopic disentanglement of surface and bulk quantum orders remains an outstanding challenge in condensed matter physics. The candidate topological superconductor PdTe has recently been proposed to host a nodeless surface gap on top of a nodal bulk state, but their direct identification and mutual coupling remained experimentally elusive. Here, we employ magnetic-field-dependent Andreev reflection spectroscopy to spectroscopically disentangle these components. At zero magnetic field, the spectra exhibit a BCS-like gap structure, consistent with dominant transport through a fully gapped surface superconducting state. Strikingly, even a weak magnetic field leads to an abrupt suppression of the Andreev-enhanced conductance (AEC), while a residual AEC, attributable to the nodal bulk state, persists to much higher magnetic fields. The transition is accompanied by pronounced magnetic hysteresis pointing to the existence of vortex dynamics at low fields. Our findings suggest that the nodal bulk gap facilitates early vortex entry, which in turn disrupts the fragile surface superconductivity. These results establish a field-tunable decoupling of surface and bulk superconductivity, and illustrate how distinct gap topologies can shape the global superconducting order in multichannel systems.

\end{abstract}

\maketitle

%\textbf{Significance statement:} Superconductors that host both topological surface states and unconventional bulk pairing provide a unique opportunity to study how distinct quantum orders coexist and interact within a single material. Yet, experimentally separating their individual roles has remained elusive. Here, we use magnetic-field-dependent Andreev reflection spectroscopy to unambiguously distinguish a nodeless, fully gapped surface superconductor from a nodal bulk condensate in PdTe. Our measurements reveal how vortex entry and nonlinear Meissner effects destabilize the surface coherence, exposing a direct connection between topology, dimensionality, and the magnetic robustness of superconductivity in complex quantum materials.

\vspace{3em}

The coexistence of multiple superconducting orders in a single material system, particularly when the corresponding superconducting gaps differ in symmetry or topology, offers an unprecedented opportunity to investigate how distinct quantum condensates interact, compete, or decouple under external stimuli. This is especially relevant in systems that host non-trivial surface states along with unconventional bulk superconductivity\cite{shang2022unconventional,linder2010unconventional,kim2018beyond,yano2023evidence,meinert2016unconventional,hossain2025superconductivity,charpentier2017induced}. In such systems dimensionality and topology intertwine to produce exotic superconductivity that is also tunable. Therefore, probing the individual response of surface and bulk superconducting channels to external perturbations, like an applied magnetic field, becomes essential for identifying the underlying pairing symmetries and understanding how topological protection gels with superconducting coherence.

PdTe, a non-centrosymmetric hexagonal chalcogenide\cite{yadav2024signature}, has recently emerged as a promising candidate in this class. Several experimental works have provided evidence for topological surface states protected by crystalline symmetries in PdTe\cite{yang2023coexistence, yadav2024signature,chapai2023evidence}. Thermodynamic and transport experiments have indicated the presence of an unconventional nodal superconducting gap in the bulk\cite{chapai2023evidence,yadav2024signature,vashist2024multigap}. Angle-resolved photoemission spectroscopy (ARPES) experiments\cite{yang2023coexistence}, suggested that PdTe simultaneously hosts a nodeless surface superconducting gap possibly of conventional $s$-wave character, and a nodal bulk gap, possibly with a mixed-parity symmetry. An unconventional order paratemeter was also suggested based on temperature dependent point contact spectroscopy experiments\cite{Das_2026}. However, a direct selective spectroscopic evidence for the coexisting components, and insight into their magnetic-field-dependent behavior, has remained lacking. A central challenge in selectively probing the surface and bulk superconducting contributions lies in the limitations of traditional probes. Surface-sensitive techniques such as ARPES and scanning tunneling spectroscopy (STS) may detect topological surface states but are not always sensitive to the site-selective superconducting orders, particularly in low-gap systems with mixed angular momentum symmetries\cite{sobota2021angle,okuda2013experimental,jiang2021topological,zhang2009experimental,noh2017experimental}. On the other hand, bulk thermodynamic measurements typically average over all superconducting channels thereby masking the individual contributions. In this context, point-contact Andreev reflection (PCAR) spectroscopy serves as a uniquely powerful tool. PCAR is capable of detecting multiple superconducting gaps through characteristic conductance spectra\cite{kovsuth2024two,gonnelli2002direct,daghero2010probing,chen2008bcs} and can access both surface and bulk components\cite{mehta2024topological}. Furthermore, the evolution of Andreev-enhanced conductance under magnetic field offers a window into how different superconducting components are suppressed, survive, or reorganize under magnetic flux entry. In this work, we have used magnetic-field-dependent PCAR spectroscopy at sub-Kelvin temperatures to probe the superconducting state of PdTe and reveal a sharp contrast in the magnetic field response of its surface and bulk superconducting condensates. %High-quality PdTe single crystals were synthesized via a melt growth technique using high-purity Pd and Te in a 1:1 molar ratio. For further details on the sample growth process, see\cite{yadav2024signature}. Electrical resistivity was measured on the as-grown crystals using a four-probe method using a Physical Property Measurement System (PPMS). The temperature dependence of the electrical resistivity ($\rho$) measured on a PdTe crystal is shown in Fig.~\ref{fig:1}(a). A sharp superconducting transition is observed at $T \sim 4.3 K$, as shown in the inset of Fig.~\ref{fig:1}(a).%At zero field, the PCAR spectra exhibit a clean, BCS-like conductance enhancement consistent with a fully gapped, nodeless superconducting state. Remarkably, even a small applied magnetic field leads to an abrupt suppression of the Andreev-enhanced conductance, while a residual reduced Andreev conductance (that is attributable to the nodal bulk gap) persists to significantly higher magnetic fields. This transition also exhibits pronounced hysteresis, strongly suggestive of vortex penetration or rearrangement at the surface. Our data suggest that the nodal structure of the bulk superconducting gap lowers the energy barrier for vortex entry, thereby facilitating early flux penetration that disrupts the fragile surface superconductivity at low fields. These results provide direct spectroscopic evidence for coexisting surface and bulk superconducting states in PdTe and demonstrate a field-driven decoupling mechanism governed by gap topology. 

\begin{figure}[h!]
    \centering
    \includegraphics[width=1\linewidth]{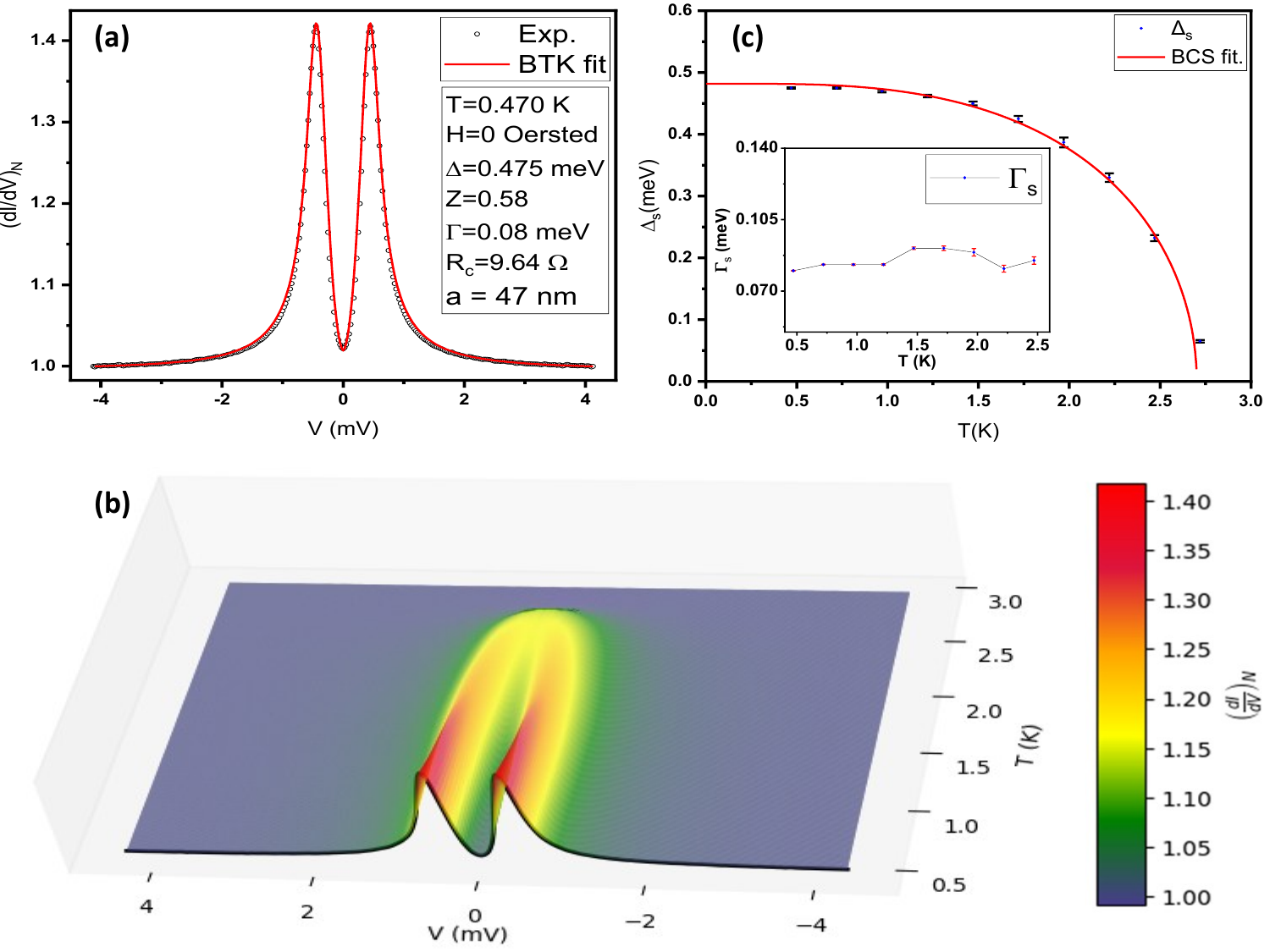}
    \caption{\textbf{PCAR spectroscopy:} (a) A representative PCAR spectrum (dots) measured at 470 mK with BTK fit (solid line). (b) Temperature evolution of the PCAR spectrum. A 2D image with single-gap BTK fits is shown in Supplementary Figure 2a. (c) $\Delta$ vs $T$ plot along with the BCS prediction (solid line). $Inset:$ Variation of $\Gamma$ with temperature, obtained from single gap BTK fit.}
    \label{fig:1}
\end{figure}

Point contact Andreev reflection (PCAR) spectroscopy measurements were performed using a home-built probe inside a He$^3$ cryostat equipped with a superconducting solenoidal magnet\cite{das2019modular}. We first characterized the low-temperature, zero-field PCAR response of PdTe. Figure~1(a) shows the normalized differential conductance $\frac{dI}{dV} (V)$ spectrum measured at $T = 0.470$~K using a ballistic Ag tip on a freshly polished PdTe surface. The spectrum exhibits a pronounced Andreev-enhanced conductance (AEC) with sharp coherence peaks near $\pm 0.48$~mV and a flat high-bias background. This flatness i.e., the absence of any additional features like critical current driven dips ensures that the contacts are in the ballistic (or diffusive) regime of mesoscopic transport, and the measured conductance is dominated by Andreev reflection\cite{sheet2004role,kumar2021nonballistic}. A modified Blonder--Tinkham--Klapwijk (BTK)\cite{blonder1982transition}, fit reproduces the lineshape with a single nodeless superconducting gap $\Delta_s = 0.475 \pm 0.005$~meV, an interface barrier parameter $Z = 0.58$, and a phenomenological broadening $\Gamma = 0.08$~meV, consistent with a highly transparent contact and minimal inelastic scattering at the interface\cite{duif1989point}. The temperature dependence of the extracted gap magnitude follows the Bardeen--Cooper--Schrieffer (BCS) prediction for an isotropic $s$ wave superconductor\cite{bardeen1957theory}. The fitted $Z$ and $\Gamma$ parameters remain nearly constant up to $T_c$, indicating the stability of the contact geometry and the absence of systematic thermal degradation. These zero-field spectra establish that the low-energy conductance is dominated by a nodeless superconducting order parameter associated with the PdTe surface states, providing a clean baseline for the subsequent field-dependent measurements.

\begin{figure}[h!]
    \centering
    \includegraphics[width=1\linewidth]{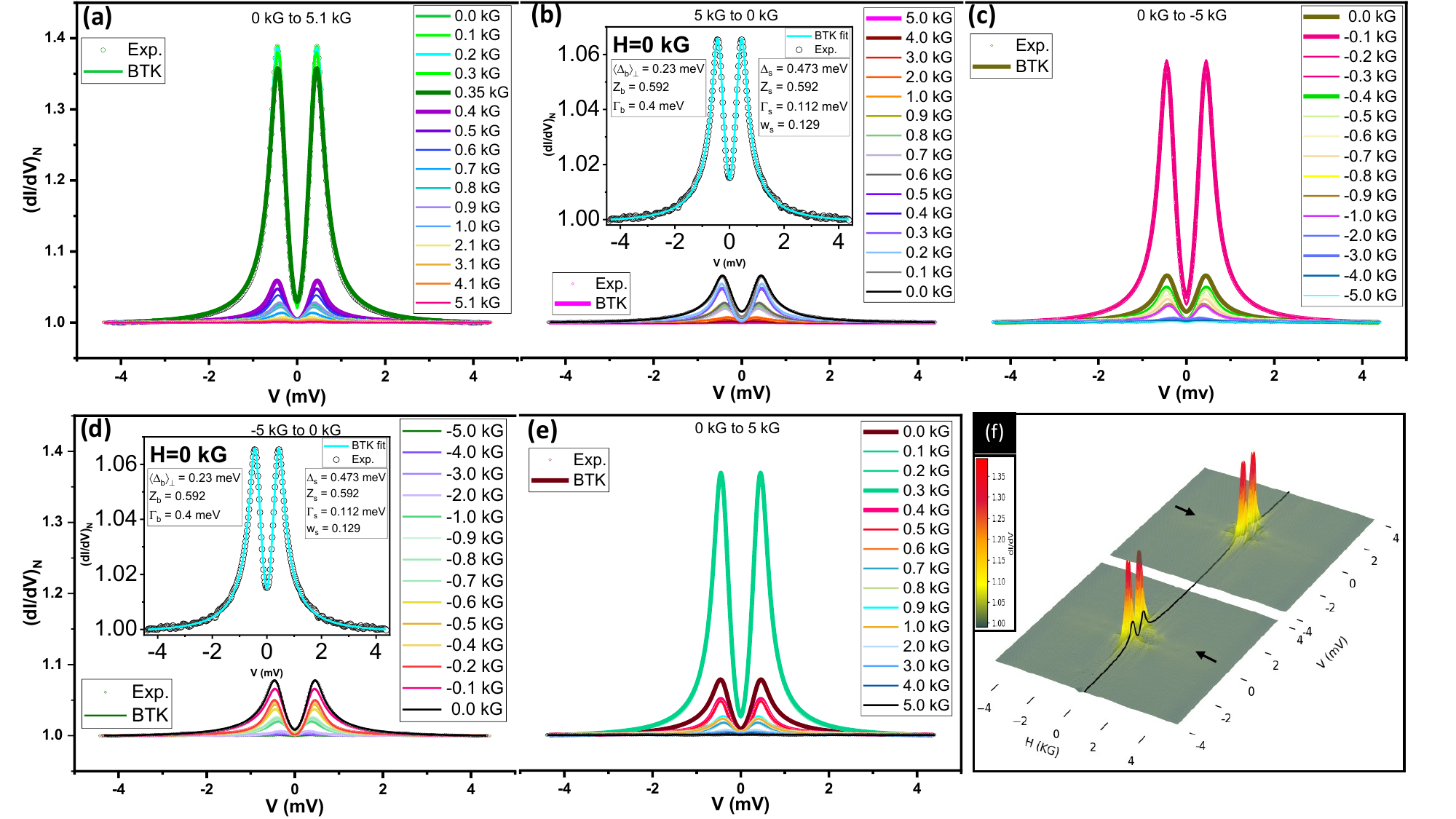}
    \caption{\textbf{Magnetic field evolution and hysteresis:} Spectra with field sweep (a) up from 0 to 5.1 kG, (b) down from 5 kG to 0, ($Inset:$ Spectrum recorded at $H=0$ kG during sweep down: OZFS not recovered.), (c) up from 0 kG to -5 kG (d) down from -5 kG to 0. ($Inset:$ Spectrum recorded at $H=0$ kG during sweep down: OZFS not recovered), up from 0 kG to 5 kG. (f)The left image shows how the spectral evolution with an applied magnetic field ramping from 5 kG to -5 kG, while the right image shows the ramp from -5 kG to 5 kG. The black line corresponds to the zero-field spectra.}
    \label{fig:2}
\end{figure}

We next investigate the evolution of the PCAR spectra under a perpendicular magnetic field, tracking both up- and down-sweep cycles. At $T = 0.47$~K and in zero applied field, the spectra exhibit pronounced Andreev enhancement of conductance (AEC) and sharp coherence peaks at $\pm 0.48$~mV. We designate this as the original zero-field spectrum (OZFS). Upon increasing $H$, the spectral shape and peak height remain essentially unchanged up to a field $H = H_{\mathrm{en}} \approx 0.35$~kG. Above this field, the AEC is abruptly suppressed, the coherence peaks broaden, and a low-energy conductance shoulder emerges. We refer to this state as the low-field suppressed spectrum (LFSS). With further field increase, both the AEC and coherence peaks are progressively diminished, and the low-bias conductance approaches the normal-state value at the critical field of the point contact, $H_{PC} \approx 4.5$~kG. To ensure complete suppression of superconductivity, the field was raised to $H = 5$~kG. This up-sweep branch is shown in Figure~2(a).  In the subsequent down-sweep [Figure~2(b)], as $H$ is lowered from 5~kG to 0~kG, the OZFS is not recovered; the spectrum remains in the LFSS state. Reversing the field polarity [Figure~2(c)], with $H$ increased from $0$ to $-5$~kG, the OZFS reappears at $H_{\mathrm{ex}}\approx$ -1~kG. Beyond $H_{\mathrm{en}} = $ -0.35~kG, the LFSS re-emerges and vanishes into the normal-state background beyond $H_{PC} =$ -4.5~kG. On the fourth and fifth branches, where $H$ is again cycled from $-5$~kG to $0$ and then $0$ to $+5$~kG, the OZFS is recovered at $H \approx +0.1$~kG. Thus, three key features are identified: (i) the OZFS, (ii) the LFSS above $H_{\mathrm{en}}$ and (iii) a pronounced hysteresis in the field dependence. The hysteresis is clearly visualized in the 3D color map of Fig.~2(f), where the black trace at $H = 0$ highlights that the fields at which the OZFS appears and disappears switch sign depending on the sweep direction.

\begin{figure}[h!]
    \centering
    \includegraphics[width=1\linewidth]{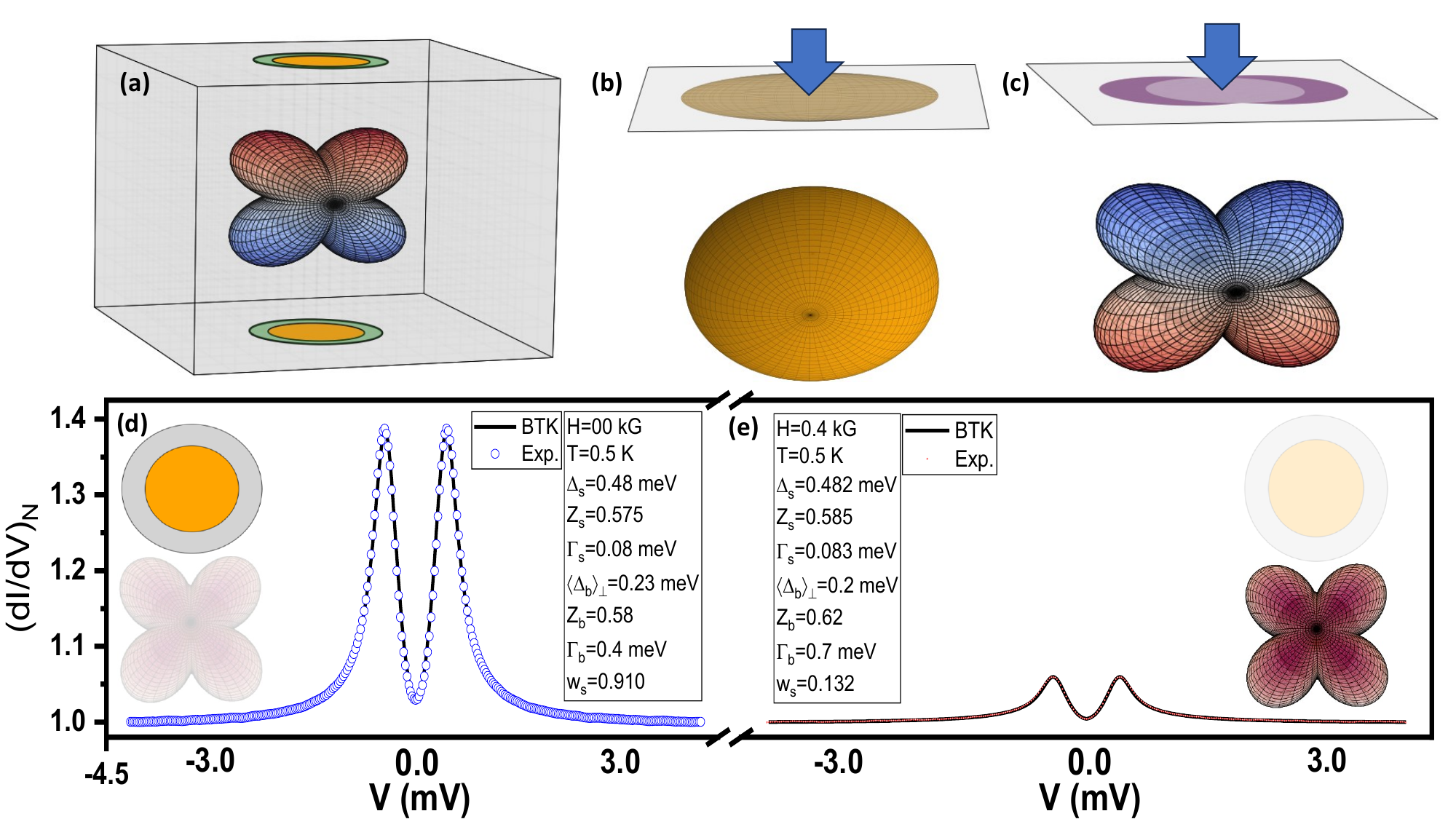}
     \caption{\textbf{Field-driven transfer of Andreev weight from surface to bulk in PdTe.} 
\textbf{(a)}~3D schematic: illustration of the bulk gap has nodal lobes (colored clover), and the surface nodeless gap. 
\textbf{(b)}~Vortex-free (Meissner) contact geometry: injected quasiparticles predominantly probe the surface channel. 
\textbf{(c)}~After first vortex entry beneath the contact: the surface response is locally suppressed and the bulk nodal channel becomes dominant. 
\textbf{(d)}~PCAR spectrum at $H=0$~G with two-gap BTK fit; the surface gap $\Delta_s$ carries most of the weight ($w_s \approx 0.91$). 
\textbf{(e)}~PCAR spectrum at $H=0.4$~kG. Andreev enhancement is quenched and the fit is bulk-dominated with $\langle\Delta_b\rangle_{\!\perp}$ with a markedly reduced surface weight ($w_s \approx 0.13$).} 
    \label{fig:3}
\end{figure}

Here we also note that the sudden conductance jump is observed most prominently in the ballistic regime, where transport across the point contact is dominated by Andreev reflection and the spectra provide a direct probe of the superconducting order parameter\cite{daghero2010probing}. For intermediate contacts, partial scattering inside the constriction broadens the spectroscopic features, so that the suppression of the surface gap at $H_\mathrm{en}$ still appears but as a smaller and less abrupt jump. In contrast, for thermal contacts, transport is dominated by local heating and resistive effects, which wash out spectroscopic signatures\cite{duif1989point}. This systematic dependence on contact regime firmly establishes that the observed phenomenon originates from the field-induced suppression of the superconducting surface gap that is probed by the point-contact current only in the ballistic (Andreev) regime of transport. The relevant data in different regimes are presented in Supplemental Information.

Before discussing the magnetic field dependent spectral behavior, we note that previous experimental works have strongly suggested a fully gapped ($s$-wave) surface superconducting state coexisting with a nodal bulk superconducting gap\cite{yang2023coexistence}. Given this background, we qualitatively interpret the field evolution of the PCAR spectra within a two-channel Andreev framework\cite{gonnelli2002direct}. Within this model, the measured (normalized) conductance is expressed as 
\begin{equation}
\begin{split}
G(V,H) &= w_{\mathrm{s}}\,\sigma_{\mathrm{s}}\bigl(eV;\Delta_{\mathrm{s}}(H),Z,\Gamma_{\mathrm{s}}(H)\bigr) \\
&\quad+ \bigl(1-w_{\mathrm{s}}\bigr)\,\sigma_{\mathrm{b}}\bigl(eV;\Delta_{\mathrm{b}}(\mathbf{k},H),Z,\Gamma_{\mathrm{b}}(H)\bigr),
\end{split}
\end{equation}
where $w_{\mathrm{s}}$ denotes the relative weight of the surface contribution, 
$Z$ is the BTK barrier strength, and $\Gamma_{\mathrm{s}}$, $\Gamma_{\mathrm{b}}$ are the quasiparticle 
broadening parameters for surface and bulk gaps, respectively. 
The functions $\sigma_{\mathrm{s}}$ and $\sigma_{\mathrm{b}}$ denote the respective normalized 
BTK conductance due to the surface and the bulk channels. 

In this complex situation where two gaps $\Delta_{\mathrm{s}}$ and $\Delta_{\mathrm{b}}$ are spatially separated, we first need to focus on which gap a transport current through a point-contact gets access to. In order to understand that, we first consider a metallic point contact injecting quasiparticles into a superconductor with interface normal (i.e., the dominant current-flow direction) $\hat{\mathbf{n}}$. In the quasiclassical description, the angle-resolved contribution to the differential conductance is weighted by the incident flux $v_{F}\cos\theta$, where $v_F$ is the Fermi velocity, and by the interface transmission $\tau(\theta)$, where $\theta$ is the angle between $\mathbf{v}_{F}$ and $\hat{\mathbf{n}}$\cite{daghero2010probing}. Within the BTK formalism\cite{blonder1982transition}, where the interface is approximated by a delta-function potential with the barrier strength given by the dimensionless parameter $Z$,
\begin{equation}
\tau(\theta)=\frac{4\cos^{2}\theta}{\bigl(4\cos^{2}\theta+Z^{2})}\!,
\qquad v_{n}\equiv v_{F}\cos\theta,
\end{equation}
and only the carrier trajectories with $v_{n}>0$ (the half–Fermi-surface $\Omega_{+}$) contribute. The measured conductance is the flux-weighted solid-angle average of the BTK kernel \cite{blonder1982transition,kashiwaya2000tunnelling}$\sigma_{\mathrm{BTK}}$ evaluated with the trajectory gap $\Delta(\hat{\mathbf{k}})$,
\begin{equation}
G(V)\;\propto\;\int_{\Omega_{+}}\! d\Omega_{\hat{\mathbf{k}}}\; 
v_{n}\,\tau(\theta)\;
\sigma_{\mathrm{BTK}}\!\bigl(eV;\,\Delta(\hat{\mathbf{k}}),Z,\Gamma_\mathrm{b},\Gamma_\mathrm{s} \bigr),
\label{eq:G-angle-average}
\end{equation}
%where $\Gamma$ is a (small) quasiparticle broadening.

Therefore, the projected (flux-weighted) gap “seen” by the injected current can be written as
\begin{equation}
\langle \Delta\rangle_{\!\perp}
\;\equiv\;
\frac{\displaystyle \int_{\Omega_{+}}\! d\Omega_{\hat{\mathbf{k}}}\;
v_{n}\,\tau(\theta)\; \bigl|\Delta(\hat{\mathbf{k}})\bigr|}
{\displaystyle \int_{\Omega_{+}}\! d\Omega_{\hat{\mathbf{k}}}\;
v_{n}\,\tau(\theta)}.
\label{eq:proj-gap}
\end{equation}
For small bias $|eV|\!\ll\!\langle \Delta\rangle_{\!\perp}$ and small $\Gamma$, the AEC is determined by this projected scale. The trajectories encountering larger $|\Delta(\hat{\mathbf{k}})|$ produce stronger Andreev reflection, while those encountering small $|\Delta(\hat{\mathbf{k}})|$ (like, near nodes) reduce the enhancement. Generally, it can be rationalized that for the zero-bias conductance: 
\begin{equation}
G(0)\;\approx\;
\mathcal{N}\;
\Big\langle \sigma_{\mathrm{BTK}}\!\bigl(0;\,|\Delta(\hat{\mathbf{k}})|,Z,\Gamma\bigr)\Big\rangle_{\!\perp}
\;\gtrsim\; 
\mathcal{N}\;\sigma_{\mathrm{BTK}}\!\bigl(0;\,\langle \Delta\rangle_{\!\perp},Z,\Gamma\bigr),
\end{equation}
where $\mathcal{N}$ is a geometric prefactor.

In the present scenario, where the surface condensate is nodeless, $\Delta_{\mathrm{s}}(\hat{\mathbf{k}})\equiv \Delta_{\mathrm{s}}$, and the bulk condensate is nodal,
$\Delta_{\mathrm{b}}(\hat{\mathbf{k}})=\Delta_{\mathrm{b},0}\,g(\hat{\mathbf{k}})$, where $g$ has zeros on lines or points of the Fermi surface and $|g|\le 1$. Their projected gaps are
\begin{equation}
\begin{split}
\langle \Delta_{\mathrm{s}}\rangle_{\!\perp} &= \Delta_{\mathrm{s}}, \\
\langle \Delta_{\mathrm{b}}\rangle_{\!\perp} &= 
\Delta_{\mathrm{b},0}\;
\frac{\displaystyle \int_{\Omega_{+}}\! d\Omega_{\hat{\mathbf{k}}}\;
v_{n}\,\tau(\theta)\; |g(\hat{\mathbf{k}})|}
{\displaystyle \int_{\Omega_{+}}\! d\Omega_{\hat{\mathbf{k}}}\;
v_{n}\,\tau(\theta)}
\;\equiv\;\Delta_{\mathrm{b},0}\,\langle |g|\rangle_{\!\perp}.
\end{split}
\end{equation}
Because $|g|$ vanishes on nodal manifolds and is $<1$ elsewhere, we have the strict inequality
\begin{equation}
0\;<\;\langle |g|\rangle_{\!\perp}\;<\;1
\quad\Rightarrow\quad
\langle \Delta_{\mathrm{b}}\rangle_{\!\perp}\;<\;\Delta_{\mathrm{b},0}.
\end{equation}

Moreover, the flux/transmission weights $v_{n}\tau(\theta)$ suppress grazing angles ($\cos\theta\!\to\!0$), so the average is dominated by near-normal trajectories; for many nodal structures (e.g.\ vertical line nodes or $d$-wave lobes), this further reduces $\langle |g|\rangle_{\!\perp}$ if nodes are intersected by those trajectories.

\begin{figure}[h!]
    \centering
    \includegraphics[width=1\linewidth]{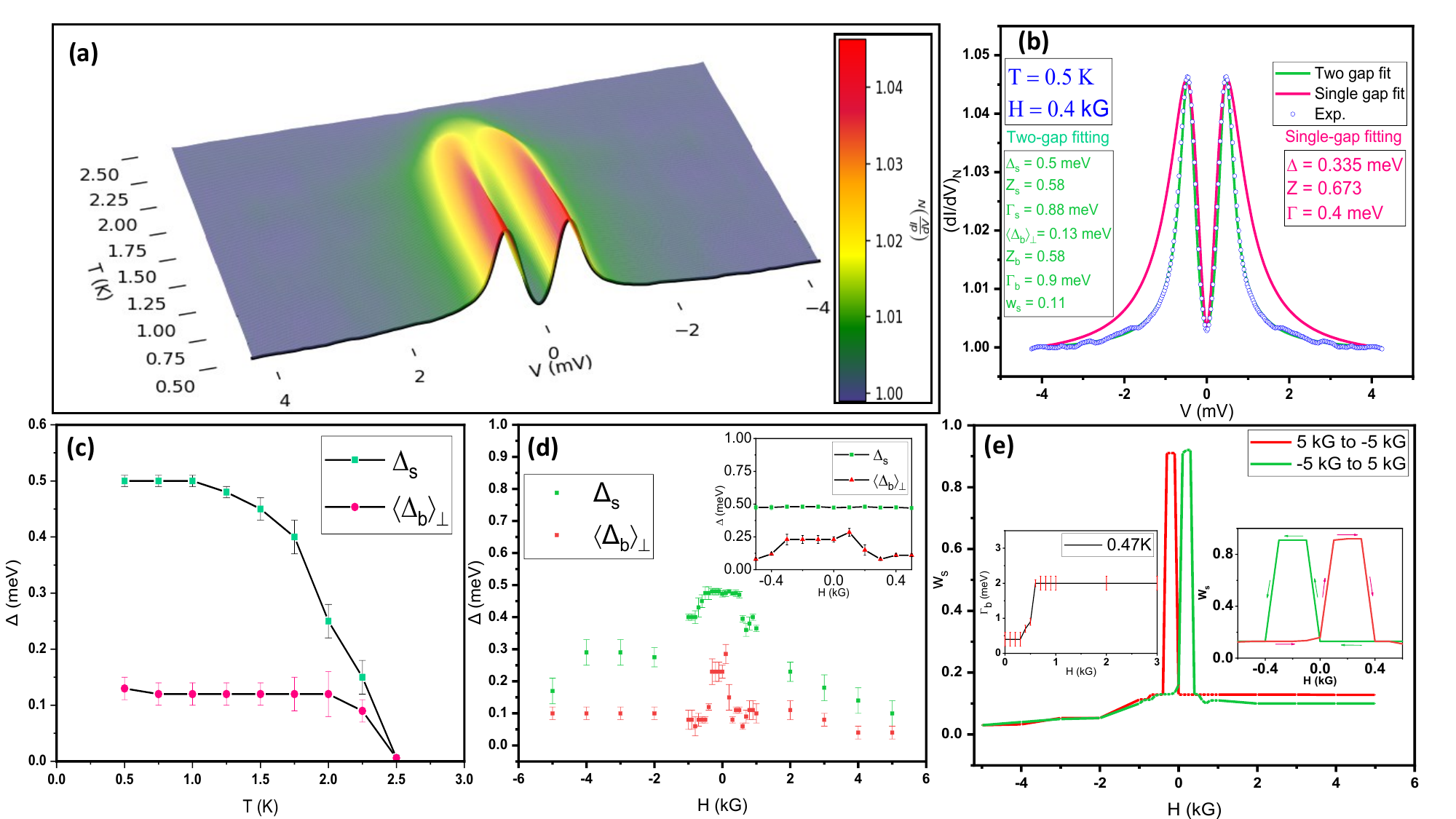}
     \caption{(a) Temperature evolution of PCAR spectra at $H=0.4$ kG. A 2D image with single-gap BTK fits is shown in Supplementary Figure 2b
(b) Spectrum at $H=0.4$ kG fitted with a two-gap BTK model (green) compared to a single-gap fit (red).  
(c) Temperature dependence of $\Delta_{\mathrm{s}}$ (green) and $\langle\Delta_{\mathrm{b}}\rangle_{\!\perp}$ (red) from two-gap fits.  
(d) Magnetic-field dependence of $\Delta_{\mathrm{s}}$ and $\langle\Delta_{\mathrm{b}}\rangle_{\!\perp}$; inset shows coherence peak variation without significant gap change in $-0.5$ to $+0.5$ kG.  
(e) Field dependence of the weight factor for $\Delta_{\mathrm{s}}$; right inset shows zoom near zero field; left inset shows the field dependence of $\Gamma_{\mathrm{b}}$.}
    \label{fig:4}
    
\end{figure}

Within the above description, the equation (1) needs to be re-formulated as below:
\begin{equation}
\begin{split}
G(V) \;=\;&\; w_{\mathrm{s}}\,
\Big\langle \sigma_{\mathrm{BTK}}\!\bigl(eV;\Delta_{\mathrm{s}},Z_{\mathrm{s}},\Gamma_{\mathrm{s}}\bigr)\Big\rangle_{\!\perp} \\
&\;+\;\bigl(1-w_{\mathrm{s}}\bigr)\,
\Big\langle \sigma_{\mathrm{BTK}}\!\bigl(eV;\Delta_{\mathrm{b},0}|g|,Z_{\mathrm{b}},\Gamma_{\mathrm{b}}\bigr)\Big\rangle_{\!\perp}
\end{split}
\end{equation}
Here it is noted that the AEC scales roughly with the corresponding projected gaps.

Now we to the discussion on the magnetic field dependent PCAR data. At zero field, the point contact predominantly probes the surface gap. This is for two reasons: a. the point-contact has significant surface sensitivity ($w_s$ is large), and b. Considering that the two gaps exist close to each other in real space and are bound to proximity be coupled, $|g| = 1$ must imply $\Delta_{b,0} = \Delta_s$. Then, as per the discussion above, 
\begin{equation}
\langle \Delta_{\mathrm{s}}\rangle_{\!\perp}=\Delta_{\mathrm{s}}
\;\;>\;\;
\langle \Delta_{\mathrm{b}}\rangle_{\!\perp}=\Delta_{\mathrm{b},0}\,\langle |g|\rangle_{\!\perp},
\end{equation}
implying
\begin{equation}
\Big\langle \sigma_{\mathrm{BTK}}\!\bigl(eV;\Delta_{\mathrm{s}},Z_{\mathrm{s}},\Gamma_{\mathrm{s}}\bigr)\Big\rangle_{\!\perp}
\;>\;
\Big\langle \sigma_{\mathrm{BTK}}\!\bigl(eV;\Delta_{\mathrm{b},0}|g|,Z_{\mathrm{b}},\Gamma_{\mathrm{b}}\bigr)\Big\rangle_{\!\perp}
\quad \text{for } |eV|\lesssim \langle \Delta_{\mathrm{b}}\rangle_{\!\perp},
\end{equation}
i.e.\ the nodeless surface channel dominates the AEC, while the nodal bulk having a smaller projected gap and finite low-energy spectral weight makes a small contribution. As a consequence, the zero field spectrum (i.e., the OZFS) is described very well within a single gap BTK formalism. This description, along with the BCS temperature dependence also confirms that $\Delta_s$ is $s$-wave.

When the magnetic field is turned on, the field- and angle-averaged conductance measured by a point contact of normal $\hat{\mathbf n}$ can be written as
\begin{align}
G(V,H) \;=\;&\; w_{\mathrm s}(H)\,
\Big\langle \sigma_{\mathrm{BTK}}\!\bigl(eV;\Delta_{\mathrm s}(H),Z_{\mathrm s},\Gamma_{\mathrm s}(H)\bigr)\Big\rangle_{\!\perp}
\nonumber \\[6pt]
&\;+\;\bigl[1-w_{\mathrm s}(H)\bigr]\,
\Big\langle \sigma_{\mathrm{BTK}}\!\bigl(eV;\Delta_{\mathrm b,0}\,|g(\hat{\mathbf k})|,Z_{\mathrm b},\Gamma_{\mathrm b}(H)\bigr)\Big\rangle_{\!\perp},
\label{eq:G-total}
\end{align}

%For $H<H_{\mathrm{en}}$ the contact region is in the clean Meissner state and hence the nodeless surface condensate is not locally pair-broken. As per London theory, the Meissner screening flow at the surface is
%\begin{equation}
%v_s(0)\;=\;\frac{e\,\mu_0\,H\,\lambda}{m^*},
%\end{equation}
%which produces a maximum Doppler shift of quasiparticle energies
%\begin{equation}
%\delta E_{\max}\;\approx\;p_F\,v_s(0)\;=\;\frac{p_F\,e\,\mu_0\,H\,\lambda}{m^*}
%\;=\;e\,\mu_0\,H\,\lambda\,v_F.
%\end{equation}
In the range $H \leq H_{en}$, the magnetic field is much smaller than the depairing field and hence the maximum Doppler shift of quasiparticle energies \cite{sakamori2024evaluation,zhu2021discovery} $\delta E_{\max}\ll \Delta_{\mathrm s}$, so the surface BTK kernel essentially remains unchanged:
\begin{equation}
w_{\mathrm s}(H)\simeq w_0,\qquad \Gamma_{\mathrm s}(H)\simeq \Gamma_{\mathrm s}(0),
\end{equation}
Hence AEC remains constant up to $H_{\mathrm{en}}$. 

Up to this point, vortex physics is also excluded from the surface. A vortex near a flat surface feels a surface barrier known as the Bean--Livingston barrier\cite{bean1964surface} that resists entry, while Meissner currents exert a Lorentz pull that tries to drag it in. As $H$ grows, the pull strengthens until it cancels the barrier at a thresold entry field. However, much before that, a nodal superconductor exhibits a nonlinear Meissner effect\cite{wilcox2022observation} where the supercurrent response is nonlinear in $v_s$ because the Doppler shift excites quasiparticles near nodes\cite{zhu2021discovery}. A known consequence of this is pre-softening of the Meissner electrodynamics which results in lowering the Bean-Livingston barrier and thus reducing the vortex entry field compared with a nodeless superconductor of the same $\lambda_0$ and $\xi_0$. In addition, the high injection current density under a point contact adds to the Meissner flow, increasing $|v_s|$ locally. Hence the local vortex entry field under the point contact can be smaller than the bulk. This explains an early OZFS$\rightarrow$LFSS transition when the halo of the first vortex thread pierces the contact region at $H = H_{en}$. %At this field two effects turn on abruptly. (i) a loss of surface-channel weight $w_{\mathrm s}$ because the vortex halo/core locally suppress the surface order parameter, and (ii) a Doppler broadening of quasiparticle energies that scales with $H$ in the mixed state.

If the contact has radius $a$ and a single vortex suppresses the surface response over an effective radius $a_{\mathrm eff}$, the fractional area removed from the surface channel by the first vortex is
\begin{equation}
f_1\;=\;\frac{a_{\mathrm eff}^{2}}{a^{2}}.
\end{equation}
For $H$ just above $H_{\mathrm{en}}$, $w_{\mathrm s}$ jumps from $w_0$ to
\begin{equation}
w_{\mathrm s}(H)\;\simeq\;w_0\,[\,1 - f_1\,],
\qquad (H\gtrsim H_{\mathrm{en}}).
\end{equation}

Therefore, the surface contribution to the total point contact current weakens abruptly. This explains the abrupt  OZFS$\rightarrow$LFSS transition. Beyond this field, the AEC primarily originates from the projection of the bulk nodal gap\cite{ramakrishnan2004evidence} $\langle\Delta_b\rangle_{\!\perp}$, along with a small contribution of $\Delta_s$. We illustrate this in Figure 3. The model nodeless surface and nodal bulk gaps are depicted in Figure 3(a). In Figure 3(b) we show how the point contact current dominantly senses the surface gap for $H < H_{en}$. The corresponding experimental spectrum is shown in Figure 3(d). Slightly above $H_{en}$, the current dominantly probes the bulk nodal gap leading to a suppressed AEC. But, the contribution of the surface still remains, but with a much weaker weight factor $w_s$. In order to verify this, we fitted the spectra obtained for $H>H_{en}$ and found that indeed such spectra cannot be described well by a single superconducting gap, but a two gap fit with a 10\% contribution of a large gap ($\Delta_s$) and a major ($\sim$~90\%) contribution of a small gap ($\langle\Delta_b\rangle_{\!\perp}$) provide an excellent fit. The BTK fitting with one gap and two distinct gaps is shown in the Figure 4(b). Further, the multigap fit also generates a significantly higher $\Gamma$ that is known to be large when the measured gap $\langle \Delta \rangle_{\!\perp}$ is an average of a large distribution of gap amplitude in the momentum space as in case of a nodal gap $\Delta_b$\cite{ramakrishnan2004evidence}. As shown in the $inset$ of Figure 4(e), $\Gamma$ shows an enhancement concurrent with the field-dependent shift of the Andreev weight factor from the surface to the bulk.  

In Figure 4(a), we present a temperature dependence of the spectrum recorded at 0.4 kG where $\Delta_s$ has been suppressed. The extracted values of the two gaps as a function of temperature are plotted in Figure 4(c). Remarkably, the supressed AEC also survives up to the same $T_c$ as that at zero field. This also indicates that the surface $s$-wave gap likely originates from proximity-induced coupling\cite{mehta2024topological} between the surface fermions and the bulk condensate. Such an inherited superconducting phase is expected to have less condensation energy, making it easier to suppress with magnetic field. In this context it is important to note that although the bulk superconducting gap in PdTe is nodal, the induced gap in the topological surface states can remain fully open. This could occur if the nodal directions of the bulk order parameter do not intersect the surface states, and the projection of the bulk pairing onto the surface yields an effectively isotropic, nodeless gap. The fragility of the surface gap can be also due to the fact that the surface condensate is quasi-2D and has a smaller effective coherence length, meaning orbital depairing from vortices and screening currents is much stronger than in the 3D bulk. This is consistent with our observation that the residual AEC completely disappears at $H_{PC} \approx$ 4.5 kG. 

Now, within the picture described above, it is straightforward to understand the hystersis effect observed in our data. Once vortices are in, getting them out is harder. As $H$ decreases, the outward Meissner drive weakens, while surface interaction and bulk pinning still hold vortices under the contact. They remain trapped near $H=0$, so $w_{\mathrm s}$ remains low and the spectrum remains LFSS-like. Reversing the field either detrap the vortex by overcoming the pinning potential or pulls antivortices that annihilate the trapped vortices. This clears the contact region and restores the surface channel, so OZFS reappears at a small opposite-sign field $H_{\mathrm{ex}}$ with $|H_{\mathrm{ex}}|< H_{\mathrm{en}}$. We have also fitted the magnetic field dependent spectra to further test the validity of our model. As shown in the $insets$ of Figure 4 (d,e), $\Delta_s$ and $w_s$ remain almost unaffected up to $H_{en}$. Above that, while $\Delta_s$ and $\langle \Delta_b \rangle_{\!\perp}$ decrease gradually, $w_s$ shows a drammatically sharp fall. In consistent with the model, $w_S$ also shows pronounced hysteresis as depicted in the $inset$ of Figure 4(e).

To conclude, our point-contact spectroscopy paints a sharp, self-consistent picture of a nodeless surface condensate coexisting with a nodal bulk condensate in PdTe. In the Meissner regime ($H<H_{\mathrm{en}}$), the Andreev-enhanced conductance (OZFS) remains intact because the Doppler shift $p_F v_s\!\propto\! H\lambda$ is $\ll \Delta_{\mathrm s}$, so the surface BTK channel dominates the projected response. At the Bean--Livingston threshold $H_{\mathrm{en}}$, a single vortex beneath the contact reduces the surface-channel weight $w_{\mathrm s}$ and facilitates the dominant measurement of the bulk nodal gap. This leads to an abrupt supression of AEC. On down-sweep, vortices remain trapped by surface interaction and pinning, preventing the recovery of OZFS at $H\!\to\!0^+$; only a small opposite field $H_{\mathrm{ex}}$ restores the vortex-free surface and the original spectrum. Our analysis shows that the full field dependence is captured by a two-channel, angle-projected BTK framework. The field-driven evolution from OZFS to LFSS, and back, turns PCAR into a directional, hysteretic probe that disentangles surface and bulk superconductivity with parameter-level precision.

A.M. and K.B. thank IISER Mohali for the Junior Research Fellowship (JRF). G.M. thanks the University Grants Commission (UGC), Government of India, for the Senior Research Fellowship (SRF). M.M. thanks the Council of Scientific and Industrial Research (CSIR), Government of India,for the Senior Research Fellowship (SRF) . We also thank Dr. Tanmoy Das and Dr. Debmalya Chakraborty for fruitful discussions. Portions of the text were edited for clarity using ChatGPT (OpenAI), and all content was reviewed and approved by the authors. G.S. acknowledges financial assistance from the Science and Engineering Research Board (SERB), Govt. of India (grant number: CRG/2021/006395). The data that support the findings of this study are available within the article and the supplementary file. The authors declare no competing financial interests.
%\bibliography{References}
\bibliographystyle{ieeetr}
\bibliography{pdte.bib}

@article{chapai2023evidence,
  title={Evidence for unconventional superconductivity and nontrivial topology in {PdTe}},
  author={Chapai, Ramakanta and Reddy, PV Sreenivasa and Xing, Lingyi and Graf, David E and Karki, Amar B and Chang, Tay-Rong and Jin, Rongying},
  journal={Scientific reports},
  volume={13},
  number={1},
  pages={6824},
  year={2023},
  publisher={Nature Publishing Group UK London}
}

@article{yadav2024signature,
  title={Signature of point nodal superconductivity in the Dirac semimetal {PdTe}},
  author={Yadav, CS and Ghosh, Sudeep Kumar and Kumar, Pankaj and Thamizhavel, A and Paulose, PL},
  journal={Physical Review B},
  volume={110},
  number={5},
  pages={054515},
  year={2024},
  publisher={APS}
}

@article{vashist2024multigap,
  title={Multigap superconductivity with non-trivial topology in a Dirac semimetal {PdTe}},
  author={Vashist, Amit and Satapathy, Bibek Ranjan and Silotia, Harsha and Singh, Yogesh and Chakraverty, S},
  journal={arXiv preprint arXiv:2408.06424},
  year={2024}
}

@article{yang2023coexistence,
  title={Coexistence of bulk-nodal and surface-nodeless Cooper pairings in a superconducting Dirac semimetal},
  author={Yang, Xian P and Zhong, Yigui and Mardanya, Sougata and Cochran, Tyler A and Chapai, Ramakanta and Mine, Akifumi and Zhang, Junyi and S{\'a}nchez-Barriga, Jaime and Cheng, Zi-Jia and Clark, Oliver J and others},
  journal={Physical review letters},
  volume={130},
  number={4},
  pages={046402},
  year={2023},
  publisher={APS}
}

@article{Das_2026,
 title={Evidence of unconventional superconductivity in Dirac semimetal PdTe via point contact spectroscopy},
 author={Das, Pritam and Dutta, Sulagna and Suman, Saurav and Vashist, Amit and Satapathy, Bibek Ranjan and Jesudasan, John and Chakraverty, Suvankar and Sensarma, Rajdeep and Raychaudhuri, Pratap},
 journal={Journal of Physics: Condensed Matter},
 volume={38},
 number={1},
 pages={015601},
 year={2025},
 month={dec},
 publisher={IOP Publishing},
}

@article{mehta2024topological,
  title={Topological surface states host superconductivity induced by the bulk condensate in {YRuB$_2$}},
  author={Mehta, Nikhlesh S and Patra, Bikash and Garg, Mona and Mohmad, Ghulam and Monish, Mohd and Bhardwaj, Pooja and Meena, PK and Motla, K and Singh, Ravi P and Singh, Bahadur and others},
  journal={Physical Review B},
  volume={109},
  number={24},
  pages={L241104},
  year={2024},
  publisher={APS}
}

@article{kovsuth2024two,
  title={Two-gap superconductivity in the noncentrosymmetric {La$_3$Se$_4$} compound},
  author={Ko{\v{s}}uth, F and Potomov{\'a}, N and Pribulov{\'a}, Z and Ka{\v{c}}mar{\v{c}}{\'\i}k, J and Naskar, M and Inosov, DS and Ash, S and Ganguli, AK and {\v{S}}olt{\`y}s, J and Cambel, V and others},
  journal={Physical Review B},
  volume={110},
  number={17},
  pages={174518},
  year={2024},
  publisher={APS}
}

@article{gonnelli2002direct,
  title={Direct Evidence for Two-Band Superconductivity in {MgB$_2$} Single Crystals  from Directional Point-Contact Spectroscopy in Magnetic Fields},
  author={Gonnelli, RS and Daghero, Dario and Ummarino, GA and Stepanov, VA and Jun, J and Kazakov, SM and Karpinski, J},
  journal={Physical review letters},
  volume={89},
  number={24},
  pages={247004},
  year={2002},
  publisher={APS}
}

@article{sheet2004role,
  title={Role of critical current on the point-contact Andreev reflection spectra between a normal metal and a superconductor},
  author={Sheet, Goutam and Mukhopadhyay, S and Raychaudhuri, Pratap},
  journal={Physical Review B},
  volume={69},
  number={13},
  pages={134507},
  year={2004},
  publisher={APS}
}

@article{kumar2021nonballistic,
  title={Nonballistic transport characteristics of superconducting point contacts},
  author={Kumar, Ritesh and Sheet, Goutam},
  journal={Physical Review B},
  volume={104},
  number={9},
  pages={094525},
  year={2021},
  publisher={APS}
}

@article{blonder1982transition,
  title={Transition from metallic to tunneling regimes in superconducting microconstrictions: Excess current, charge imbalance, and supercurrent conversion},
  author={Blonder, GE and Tinkham, m M and Klapwijk, TM},
  journal={Physical Review B},
  volume={25},
  number={7},
  pages={4515},
  year={1982},
  publisher={APS}
}

@article{bardeen1957theory,
  title={Theory of superconductivity},
  author={Bardeen, John and Cooper, Leon N and Schrieffer, John Robert},
  journal={Physical review},
  volume={108},
  number={5},
  pages={1175},
  year={1957},
  publisher={APS}
}

@article{sakamori2024evaluation,
  title={Evaluation of Doppler shifts in d-wave superconductor tunneling junctions},
  author={Sakamori, Takashi and Matsuoka, Kenki and Teshigawara, Mitsuhiro and Mawatari, Yasunori and Yada, Keiji and Tanaka, Yukio and Kashiwaya, Satoshi},
  journal={Physical Review B},
  volume={110},
  number={22},
  pages={224503},
  year={2024},
  publisher={APS}
}

@article{zhu2021discovery,
  title={Discovery of segmented Fermi surface induced by Cooper pair momentum},
  author={Zhu, Zhen and Papaj, Micha{\l} and Nie, Xiao-Ang and Xu, Hao-Ke and Gu, Yi-Sheng and Yang, Xu and Guan, Dandan and Wang, Shiyong and Li, Yaoyi and Liu, Canhua and others},
  journal={Science},
  volume={374},
  number={6573},
  pages={1381--1385},
  year={2021},
  publisher={American Association for the Advancement of Science}
}

@article{bean1964surface,
  title={Surface barrier in type-II superconductors},
  author={Bean, CP and Livingston, JD},
  journal={Physical Review Letters},
  volume={12},
  number={1},
  pages={14},
  year={1964},
  publisher={APS}
}

@article{wilcox2022observation,
  title={Observation of the non-linear Meissner effect},
  author={Wilcox, JA and Grant, MJ and Malone, L and Putzke, C and Kaczorowski, D and Wolf, T and Hardy, F and Meingast, C and Analytis, JG and Chu, J-H and others},
  journal={Nature communications},
  volume={13},
  number={1},
  pages={1201},
  year={2022},
  publisher={Nature Publishing Group UK London}
}

@article{sobota2021angle,
  title={Angle-resolved photoemission studies of quantum materials},
  author={Sobota, Jonathan A and He, Yu and Shen, Zhi-Xun},
  journal={Reviews of Modern Physics},
  volume={93},
  number={2},
  pages={025006},
  year={2021},
  publisher={APS}
}

@article{zhang2009experimental,
  title={Experimental Demonstration of Topological Surface States Protected by Time-Reversal Symmetry},
  author={Zhang, Tong and Cheng, Peng and Chen, Xi and Jia, Jin-Feng and Ma, Xucun and He, Ke and Wang, Lili and Zhang, Haijun and Dai, Xi and Fang, Zhong and others},
  journal={Physical Review Letters},
  volume={103},
  number={26},
  pages={266803},
  year={2009},
  publisher={APS}
}

@article{okuda2013experimental,
  title={Experimental evidence of hidden topological surface states in {PbBi$_4$Te$_7$}},
  author={Okuda, Taichi and Maegawa, Takamasa and Ye, Mao and Shirai, Kaito and Warashina, Takuya and Miyamoto, Koji and Kuroda, Kenta and Arita, Masashi and Aliev, Ziya S and Amiraslanov, Imamaddin R and others},
  journal={Physical review letters},
  volume={111},
  number={20},
  pages={206803},
  year={2013},
  publisher={APS}
}

@article{jiang2021topological,
  title={Topological surface states in superconducting {CaBi$_2$}},
  author={Jiang, Qi and Wang, Deyang and Liu, Zhengtai and Jiang, Zhicheng and Qian, Haoji and Shen, Xiaoping and Li, Ang and Shen, Dawei and Qiao, Shan and Ye, Mao},
  journal={Physical Review B},
  volume={104},
  number={24},
  pages={245112},
  year={2021},
  publisher={APS}
}

@article{daghero2010probing,
  title={Probing multiband superconductivity by point-contact spectroscopy},
  author={Daghero, Dario and Gonnelli, RS},
  journal={Superconductor Science and Technology},
  volume={23},
  number={4},
  pages={043001},
  year={2010},
  publisher={IOP Publishing}
}

@article{noh2017experimental,
  title={Experimental realization of type-II Dirac fermions in a {PdTe$_2$} superconductor},
  author={Noh, Han-Jin and Jeong, Jinwon and Cho, En-Jin and Kim, Kyoo and Min, BI and Park, Byeong-Gyu},
  journal={Physical review letters},
  volume={119},
  number={1},
  pages={016401},
  year={2017},
  publisher={APS}
}

@article{shang2022unconventional,
  title={Unconventional superconductivity in topological Kramers nodal-line semimetals},
  author={Shang, Tian and Zhao, Jianzhou and Hu, Lun-Hui and Ma, Junzhang and Gawryluk, Dariusz Jakub and Zhu, Xiaoyan and Zhang, Hui and Zhen, Zhixuan and Yu, Bocheng and Xu, Yang and others},
  journal={Science Advances},
  volume={8},
  number={43},
  pages={eabq6589},
  year={2022},
  publisher={American Association for the Advancement of Science}
}

@article{linder2010unconventional,
  title={Unconventional superconductivity on a topological insulator},
  author={Linder, Jacob and Tanaka, Yukio and Yokoyama, Takehito and Sudb{\o}, Asle and Nagaosa, Naoto},
  journal={Physical review letters},
  volume={104},
  number={6},
  pages={067001},
  year={2010},
  publisher={APS}
}

@article{yano2023evidence,
  title={Evidence of unconventional superconductivity on the surface of the nodal semimetal {CaAg$_{1- x}$Pd$_x$P}},
  author={Yano, Rikizo and Nagasaka, Shota and Matsubara, Naoki and Saigusa, Kazushige and Tanda, Tsuyoshi and Ito, Seiichiro and Yamakage, Ai and Okamoto, Yoshihiko and Takenaka, Koshi and Kashiwaya, Satoshi},
  journal={Nature Communications},
  volume={14},
  number={1},
  pages={6817},
  year={2023},
  publisher={Nature Publishing Group UK London}
}

@article{charpentier2017induced,
  title={Induced unconventional superconductivity on the surface states of ${Bi_2Te_3}$ topological insulator},
  author={Charpentier, Sophie and Galletti, Luca and Kunakova, Gunta and Arpaia, Riccardo and Song, Yuxin and Baghdadi, Reza and Wang, Shu Min and Kalaboukhov, Alexei and Olsson, Eva and Tafuri, Francesco and others},
  journal={Nature communications},
  volume={8},
  number={1},
  pages={2019},
  year={2017},
  publisher={Nature Publishing Group UK London}
}

@article{meinert2016unconventional,
  title={Unconventional superconductivity in ${YPtBi}$ and related topological semimetals},
  author={Meinert, Markus},
  journal={Physical review letters},
  volume={116},
  number={13},
  pages={137001},
  year={2016},
  publisher={APS}
}

@article{hossain2025superconductivity,
  title={Superconductivity and a van Hove singularity confined to the surface of a topological semimetal},
  author={Hossain, Md Shafayat and Islam, Rajibul and Cheng, Zi-Jia and Muhammad, Zahir and Zhang, Qi and Guguchia, Zurab and Krieger, Jonas A and Casas, Brian and Jiang, Yu-Xiao and Litskevich, Maksim and others},
  journal={Nature Communications},
  volume={16},
  number={1},
  pages={3998},
  year={2025},
  publisher={Nature Publishing Group UK London}
}

@article{kim2018beyond,
  title={Beyond triplet: Unconventional superconductivity in a spin-3/2 topological semimetal},
  author={Kim, Hyunsoo and Wang, Kefeng and Nakajima, Yasuyuki and Hu, Rongwei and Ziemak, Steven and Syers, Paul and Wang, Limin and Hodovanets, Halyna and Denlinger, Jonathan D and Brydon, Philip MR and others},
  journal={Science advances},
  volume={4},
  number={4},
  pages={eaao4513},
  year={2018},
  publisher={American Association for the Advancement of Science}
}

@article{das2019modular,
  title={A modular point contact spectroscopy probe for sub-Kelvin applications},
  author={Das, Shekhar and Sheet, Goutam},
  journal={Review of Scientific Instruments},
  volume={90},
  number={10},
  year={2019},
  publisher={AIP Publishing}
}

@article{duif1989point,
  title={Point-contact spectroscopy},
  author={Duif, AM and Jansen, AGM and Wyder, P},
  journal={Journal of Physics: Condensed Matter},
  volume={1},
  number={20},
  pages={3157},
  year={1989},
  publisher={IOP Publishing}
}

@article{ramakrishnan2004evidence,
  title={Evidence of Gap Anisotropy in Superconducting {YNi$_2$B$_2$C} Using Directional Point-Contact Spectroscopy},
  author={Ramakrishnan, S and Sheet, Goutam and Jaiswal-Nagar, D and Takeya, Hiroyuki and Raychaudhuri, Pratap},
  journal={Physical Review Letters},
  volume={93},
  year={2004},
  publisher={American Physical Society (APS)}
}

@article{kashiwaya2000tunnelling,
  title={Tunnelling effects on surface bound states inunconventional superconductors},
  author={Kashiwaya, Satoshi and Tanaka, Yukio},
  journal={Reports on Progress in Physics},
  volume={63},
  number={10},
  pages={1641},
  year={2000},
  publisher={IOP Publishing}
}

@article{chen2008bcs,
  title={A BCS-like gap in the superconductor {SmFeAsO$_0.85$F$_0. 15$}},
  author={Chen, TY and Tesanovic, Z and Liu, RH and Chen, XH and Chien, CL},
  journal={Nature},
  volume={453},
  number={7199},
  pages={1224--1227},
  year={2008},
  publisher={Nature Publishing Group UK London}
}
\end{document}